# Perpendicular magnetic tunnel junctions with multi-interface free layer

Pravin Khanal[1], Bowei Zhou[1], Magda Andrade[1], Yanliu Dang[2,3], Albert Davydov[2], Ali Habiboglu[1], Jonah Saidian[1], Adam Laurie[1], Jian-Ping Wang[4], Daniel B Gopman[2] and Weigang Wang[1]*

1. Department of Physics, University of Arizona, Tucson, AZ 85721, USA
2. Materials Science & Engineering Division, National Institute of Standards and Technology, Gaithersburg, MD 20899, USA
3. Department of Electrical & Computer Engineering, Purdue University, West Lafayette, Indiana 47907, USA
4. Department of Electrical & Computer Engineering, University of Minnesota, Minneapolis, MN 55455, USA

Future generations of magnetic random access memory demand magnetic tunnel junctions that can provide simultaneously high magnetoresistance, strong retention, low switching energy and small cell size below 10nm. Here we study perpendicular magnetic tunnel junctions with composite free layers where multiple ferromagnet/nonmagnet interfaces can contribute to the thermal stability. Different nonmagnetic materials (MgO, Ta, Mo) have been employed as the coupling layers in these multi-interface free layers. The evolution of junction properties under different annealing conditions is investigated. A strong dependence of tunneling magnetoresistance on the thickness of the first CoFeB layer has been observed. In junctions where Mo and MgO are used as coupling layers, large tunneling magnetoresistance above 200% has been achieved after 400°C annealing.

*wgwang@physics.arizona.edu

Magnetic tunnel junction with perpendicular magnetic anisotropy (pMTJ) is one of the leading candidates for non-volatile magnetic random-access memories (MRAM).[1,2] Ideally, MRAM cells made of pMTJ should exhibit high tunneling magnetoresistance (TMR >200%), be thermally stable at room temperature (>10 years), occupy only a small footprint (< 10 nm), and operate with minimum energy consumption by spin-transfer torques (STT),[3,4] spin orbit torques (SOT),[5] voltage controlled magnetic anisotropy (VCMA),[6,7] or other methods.[8–10] In particular when the recording layer of a pMTJ is a single ferromagnetic (FM) layer with interfacial PMA, the areal perpendicular energy density is usually 1-2 mJ/m$^2$, which cannot provide enough retention when the junction size is below 10 nm.[11–14] Generally, three types of pMTJ are under investigation to solve this problem. In the first type, FMs with bulk perpendicular magnetic anisotropy, such as FePd,[15,16] or MnGa,[17] alloys are employed, where the thermal stability factor ($\Delta$) can be increased by increasing the thickness of the FM layer. However, the TMR in these junctions are typically lower than that of CoFeB/MgO due to the lack of coherent tunneling effect.[18] In the second approach, shape anisotropy is employed to promote the out-of-plane easy axis in junctions where the thickness of the CoFeB free layer is much larger than the lateral dimension of the junction.[19] A pMTJ smaller than 3 nm has been successfully achieved with this approach and STT switching has been demonstrated.[20] This method, however, requires a thick free layer, which may lead to difficulties in obtaining fast switching and device fabrication. In the third method, multiple CoFeB/non-magnet(NM) interfaces are used to enhance the overall PMA of the free layer, where both the CoFeB and NM are limited to very thin thickness (~1nm). Significant increases of $\Delta$ and switching efficiency have been realized when the single CoFeB free layer was replaced by a CoFeB/NM/CoFeB/MgO composite free layer.[21–24] Further modification of stack structure was used to increase the coupling and thermal stability of the free layer.[25,26] For example, the performance of the pMTJs with a quad-interface free layer has been shown to be substantially enhanced compared to that of double-interface.[27]

In this work, we investigated the transport properties in pMTJs with multi-interface free layers (MIFL), where different NM materials, such as MgO, Ta, and Mo, have been employed as the coupling layers to enhance the TMR. The perpendicular magnetic anisotropy, interlayer magnetic coupling and magnetoresistance can be controlled by varying the FM thickness, NM thickness and post-growth thermal annealing treatment. As a result, a TMR ratio as high as 212% have been achieved in junctions with MIFLs where three CoFeB layers are coupled through Mo and MgO, which to the best of our knowledge is the highest TMR in this type of pMTJs.

The films in this work were fabricated in a 12-source UHV sputtering system (AJA International) with a base pressure of 10$^{-7}$ Pa (10$^{-9}$ Torr). The structure of the MTJ films is Si/SiO$_2$/Ta(12 nm)/Ru(15 nm)/Ta(10 nm)/Mo (0.75 nm)/Co$_{20}$Fe$_{60}$B$_{20}$(1 nm)/MgO(1.5-3.5 nm)/MIFL/Mo (1.9 nm)/Ta(10 nm)/Ru(20 nm). Different MIFL film compositions have been synthesized as detailed below. Circular pMTJs with diameters ranging from 2 μm to 100 μm were patterned by conventional microfabrication process involving photolithography and ion beam etching, and subsequently annealed under varying conditions to be described below. Detailed information on sample fabrication and characterization can be found in our previous publications.[8,12,14,28,29]

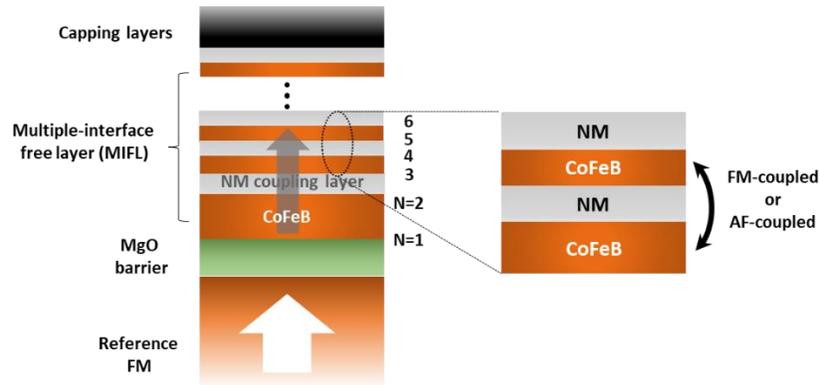

Figure 1. Schematic representation of a pMTJ with multiple-interface free layer (MIFL) where a number of FM layers are coupled through NM layers to function as a single magnetic layer

The schematic structure of a pMTJ with MIFL is shown in Figure 1. Multiple FM layers are coupled through the NM layers to behave like a single free layer of the pMTJ. When FM layers are thin (≈1 nm), the hybridization of 3$d$ orbitals of FM with the 2$p$ orbitals of O,[30] or with the 5$d$ orbitals of heavy metals,[31,32] leads to the interfacial PMA. When properly designed, the PMA from each FM/NM interface can add up to each other therefore providing a sufficiently large $\Delta$ for small pMTJ cells. The coupling between the FM layers can be either FM or AF, depending on the needs of a particular application, which can be controlled by the materials and thickness of the FM and NM layers, as well as post-fabrication processes such as thermal annealing. AF-coupled FMs may provide technologically superior performance as it has been predicted that the switching speed can be increased in AF-coupled FMs.[33] This structure is similar to those used in perpendicular spin valves,[34] and pMTJ,[35] except now it combines FM with high spin-polarization with multiple FM/NM interfaces that can provide at the same time a large TMR and strong retention after high temperature annealing at 400°C or above. CoFeB alloys are a promising candidate for the first FM in the MIFL, owing to the symmetry-conserved tunneling effect at the Fe/MgO interface.[18] However, for the other layers in the MIFL, a combination of different materials could be used to maximize the PMA and interlayer coupling. Ideally, for any pMTJ at a given lateral size, the number of FM layers and therefore the total number of interfaces ($N$) contributing to PMA energy density could be increased until the desired $\Delta$ is reached.

We first investigated MIFLs with MgO as the coupling layer. In addition to the larger TMR, the CoFeB/MgO interface also provides a strong interfacial PMA.[36,37] Interlayer exchange coupling in epitaxial Fe/MgO structures has been observed previously, where AF coupling has been observed when the MgO thickness is less than 0.8nm.[38,39] However, most of these studies were performed with thick Fe layers where the magnetic easy axis lies within the plane. For sputtered MgO that is sandwiched between two perpendicularly magnetized Co layers, AF coupling was also found even when MgO was as thick as 1.3 nm.[40] For the MIFL depicted in Figure 1, it is desirable to have a MgO coupling layer that is thick enough to support a strong coupling, but thin enough to contribute only minimal additional series resistance to the overall resistance of the pMTJ – which may be satisfied if current can conduct across pinholes within the thin MgO layer. The magnetic properties of the MIFL with MgO was first investigated in a sample with the structure of [CoFeB(0.75 nm)/MgO(0.8 nm)]$_3$ by a vibrating sample magnetometer (VSM). The film exhibits PMA as shown in Figure 2a, where an in-plane anisotropy field larger than 1 T (10 kOe) can be observed. The MH loops in the low field region is shown in Figure 2b, where the remanent magnetic moment is about 20 nA·m$^2$ (20 μemu) which is about one-third of the saturation moment (60 nA·m$^2$), indicating the three CoFeB layers are AF-coupled.

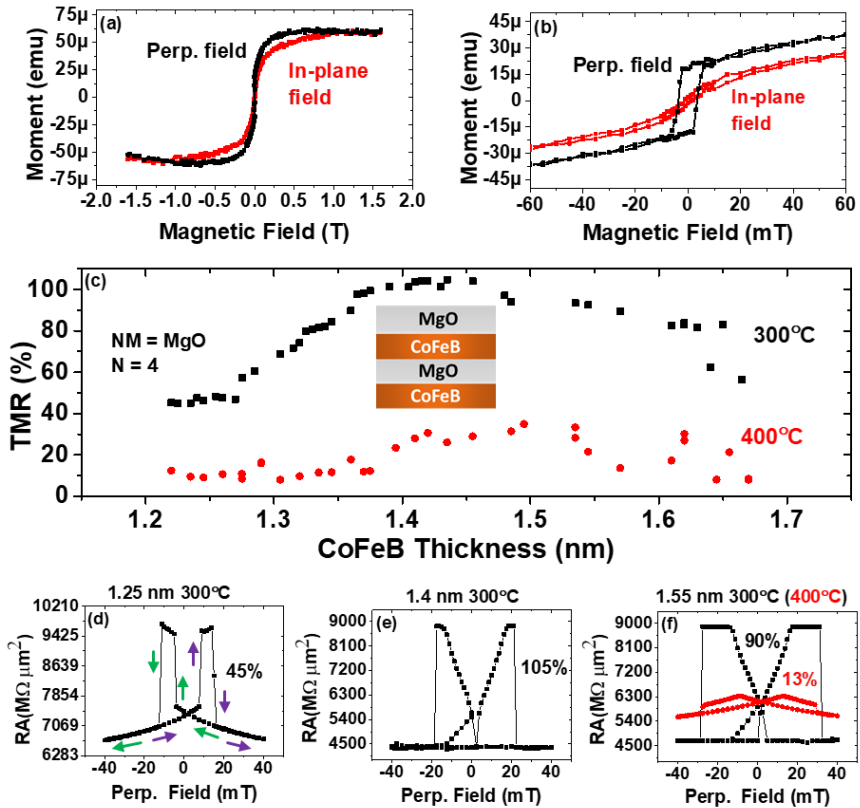

Figure 2. (a) Hysteresis loops of [CoFeB(0.75 nm)/MgO(0.8 nm)]$_{x3}$ measured under different magnetic fields. (b) The same curves at the low-field region. (c) CoFeB thickness dependence of the TMR in pMTJs with MgO-MIFLs (Inset: schematic of MgO-MIFL). The samples were first annealed at 300°C for 10 min, then 400°C for another 10 min. (d-f) representative TMR curves of the pMTJs after the 300°C annealing. The arrows in (d) show the representative magnetic field sweeping direction [purple (green) towards the positive(negative) field direction]. The red curve in (f) is the TMR of the same sample after the 400°C annealing.

Next, pMTJs with MIFLs of the structure of CoFeB (1.2 nm - 1.7 nm)/ MgO (0.9 nm)/CoFeB (1.3 nm) were fabricated. These pMTJs are denoted as *N*=4 because the total PMA originates from three CoFeB/MgO interfaces and one CoFeB/Mo interface (recall the capping layers are Mo/Ta/Ru). Since the first CoFeB in the MIFL is the one contributing to both TMR and PMA, it is critical to study the thickness dependence of this layer. The TMR of these junctions after annealing at 300°C for 10 min is plotted in Figure 2c. The TMR is about 45% when CoFeB is 1.25 nm thick (Figure 2d), which increases to more than 100% when CoFeB is 1.45 nm thick. Further increase of CoFeB thickness beyond this point leads to a slight decrease of TMR (Figure 2f). Note MTJs located on the edge of the wafer were not included due to poor pattering as a result of edge bead formation. The TMR is, however, reduced across the entire thickness series when the same pMTJs were annealed again (after testing at RT)at 400°C for 10 min. A number of processes simultaneously occur during the annealing process, most importantly the formation of the CoFe(001)/MgO(001) epitaxial structure with the B diffusing out of CoFeB, and the reduction of interfacial oxidation which eventually leads to proper hybridization of Oxygen 2$p$ orbitals and Fe/Co 3$d$ orbitals that is required for a strong PMA. Typically the parallel state resistance ($R_P$) of the junction momentarily drops at the beginning of the annealing, resulting from the initial establishment of the highly conductive $\Delta_1$ channel, followed by a steady increase due to the gradual deterioration of that channel when other atomic species inevitably diffuse into the barrier.[41,42] Despite the increase of $R_P$, the TMR may continue to increase at 400°C for up to a few hours of annealing, provided that increases in the anti-parallel state resistance ($R_{AP}$) due to the reduction of the $\Delta_2$ and $\Delta_5$ conduction channels outpaces of the increase of $R_P$.[41,43] The comparison of the TMR curves from the same junction after annealing at 300°C and 400°C is presented in Figure 2f. In addition to the increase of $R_P$, we note that R$_{AP}$ is decreased as shown by the red TMR curve. The decrease of $R_{AP}$ in Figure 2f is accompanied by the disappearance of the sharp switching in the TMR curve, which is likely due to the reduction of PMA of the MIFL, instead of a more transport-intrinsic reason that is usually only expected when the annealing is much longer.[41] The reduction of PMA might be a result of Boron aggregation at the CoFeB/MgO interface in the absence of any "Boron-absorbing" layer adjacent to the CoFeB layer.

Next, we investigated pMTJs with the MIFL of the structure of CoFeB (1.2 nm-1.7 nm)/ Ta(1 nm)/CoFeB (1 nm)/MgO (0.8 nm), which is denoted by $N$=4 with Ta in Figure 3 since now there are two CoFeB/MgO interfaces and two CoFeB/Ta interfaces that each contribute to PMA. This structure is similar to what was used previously,[21–24] except here we chose the Ta at 1nm. Maximum AF-coupling was observed in Co/Ta superlattices when Ta is near 0.7nm.[44] For CoFeB/Ta/CoFeB, a sizable AF coupling was obtained when the thickness of Ta is around 1nm,[45] which is in agreement with our VSM results shown in Figures 3a and 3b. After the 300°C annealing for 10 min, maximal TMR of 135 % was obtained as shown in Figure 3c, which is considerably better in those shown in Figure 2. Reasonably high TMR (> 100 %) is present in pMTJs across a wide range of CoFeB thickness. A representative TMR curve under this annealing condition is shown in Figure 3d, with sharp transitions between states and a clear AP state. This improved TMR behavior is likely related to the fact that the Ta coupling layer more readily absorbs Boron diffusing out from the CoFeB layer, compared to a MgO coupling layer.

Subsequent annealing of these pMTJs at 400°C for 10 min substantially increased the maximum TMR to above 180 % as shown in Figure 3f, which is noticeably higher than previous reports.[21–24] However, the TMR starts to fall off after the CoFeB thickness exceeds 1.5 nm and exhibits large fluctuations when CoFeB is thicker than 1.6nm. This reduction in the TMR is attributed to the loss of the AP state as shown in Figure 3g. Usually, the strength and sign of the interlayer coupling is sensitively depended on the thickness of the FM and NM layers.[40,46] Here another complexity is involved, which is the PMA of MIFL. Due to the relatively small formation energy between Ta and Fe, it is known that the PMA of MgO/CoFeB/Ta is not stable when annealed at 400°C,[12] which leads to the deterioration of PMA of the MIFL stack. Further annealing of these pMTJs at 500°C leads to a dramatic reduction of TMR to nearly zero, consistent with previous studies.[12]

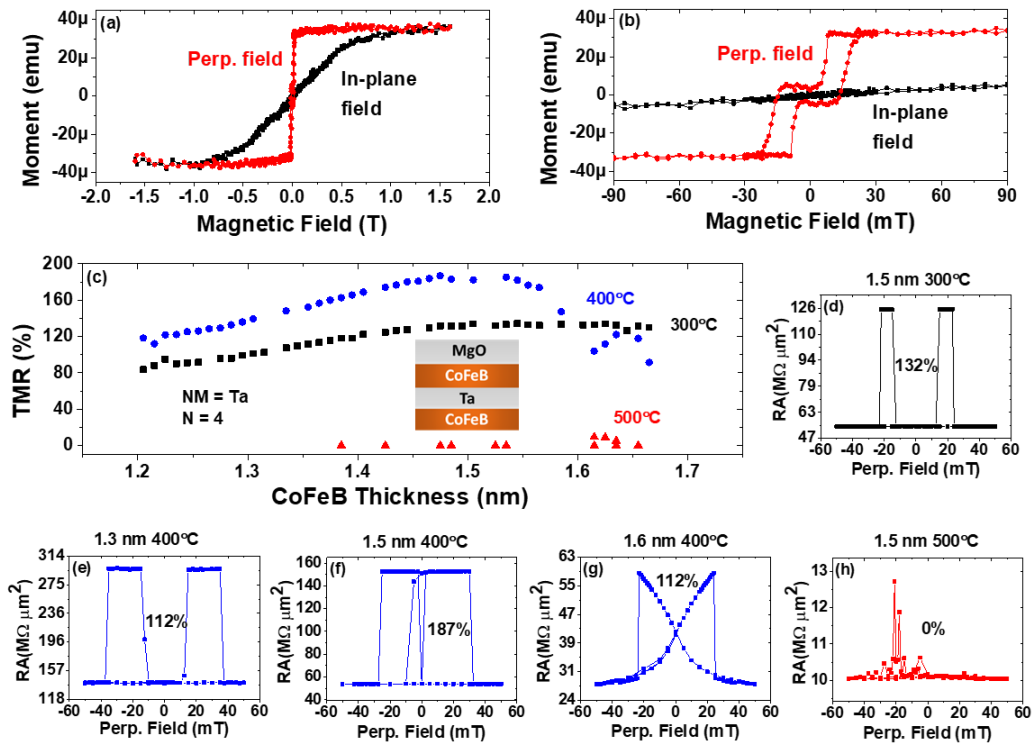

Figure 3. (a) Hysteresis loops of MgO/ CoFeB/Ta(1nm)/CoFeB/MgO measured on different magnetic fields. (b) The same curves at low field-region. (c) CoFeB thickness dependence of the TMR in pMTJs with Ta-MIFLs (Inset: schematic of Ta-MIFL). The samples were successive annealed at each temperature for 10 min. (d) TMR curve of the 1.5 nm sample after the 300°C annealing. (e-g) TMR curves of three pMTJs after the 400°C annealing. (h) TMR of the 1.5 nm sample after the 500°C annealing.

It has been shown that pMTJs with Mo as the heavy metal layer exhibited much higher TMR than that of Ta[12,14,47]. Therefore, MIFLs with Mo as the coupling layer may provide larger TMR as well. Another benefit of Mo is that its interlayer exchange coupling energy is larger compared to that of Ta.[44] It was also shown that Mo can substantially enhance damping.[48] pMTJs with three CoFeB layers in the MIFL have been fabricated. The MIFL stack structure is CoFeB (1.2 nm-1.7 nm)/Mo (0.9 nm)/ CoFeB (1 nm)/MgO(0.9 nm)/ CoFeB (1.3 nm). These samples are denoted as *N*=6 with Mo as plotted in Figure 4. The VSM results show the strong AF coupling for Mo(0.9nm) are presented in Figures 4a and 4b. The TMR ratios of the samples after annealing at 300°C for 10 min (red curve) are presented in Figure 4c. The maximum TMR in pMTJs with Mo-MIFL is similar to that of Ta-MIFL under this annealing condition. However, the TMR starts to decay in pMTJs with Mo-MIFL when CoFeB is thicker than 1.5 nm. By comparison, the TMR with Ta-MIFL gains a slight increase over 1.5 nm to 1.6 nm under the same annealing condition as shown in Figure 3c. This feature of Mo-MIFL becomes more pronounced after the annealing at 400°C for 10 min as shown in the black curve of Figure 4c, where TMR drops sharply when CoFeB is thicker than 1.4 nm. When the pMTJs are annealed at 400°C for one hour, the overall TMR has been increased (blue curve), with maximum TMR reaching 212 % as shown in Figure 4e, which is even higher than the TMR in pMTJs we obtained previously with a single CoFeB layer as the free layer.[14] However, TMR quickly drops when CoFeB thickness exceeds 1.42 nm. The reduction of TMR is again related to the disappearance of the AP state as shown in the Figure 4f, which is likely due to the loss of PMA of the MIFL. These results highlight the very sensitive dependence of TMR on the first CoFeB layer thickness in the MIFL. The AF coupling peak with Mo varies in different reports, ranging from 0.5 nm,[44] to 0.8 nm.[49] In another series of pMTJ where the Mo in the MIFL is slightly thicker (1 nm), a more pronounced AF coupling of the free layer can be seen as shown in the inset of Figure 4c. The free layer switching fields are obviously not symmetric about the zero magnetic field, which is a signature of the AF coupling of the CoFeB in the MIFL. The behavior of these pMTJs is plotted in the green curve in Figure 4c. Interestingly, the range of CoFeB thicknesses where high TMR ratio is observed is extended by nearly 0.1 nm, as evident from the shift of the green curve relative to the black one.

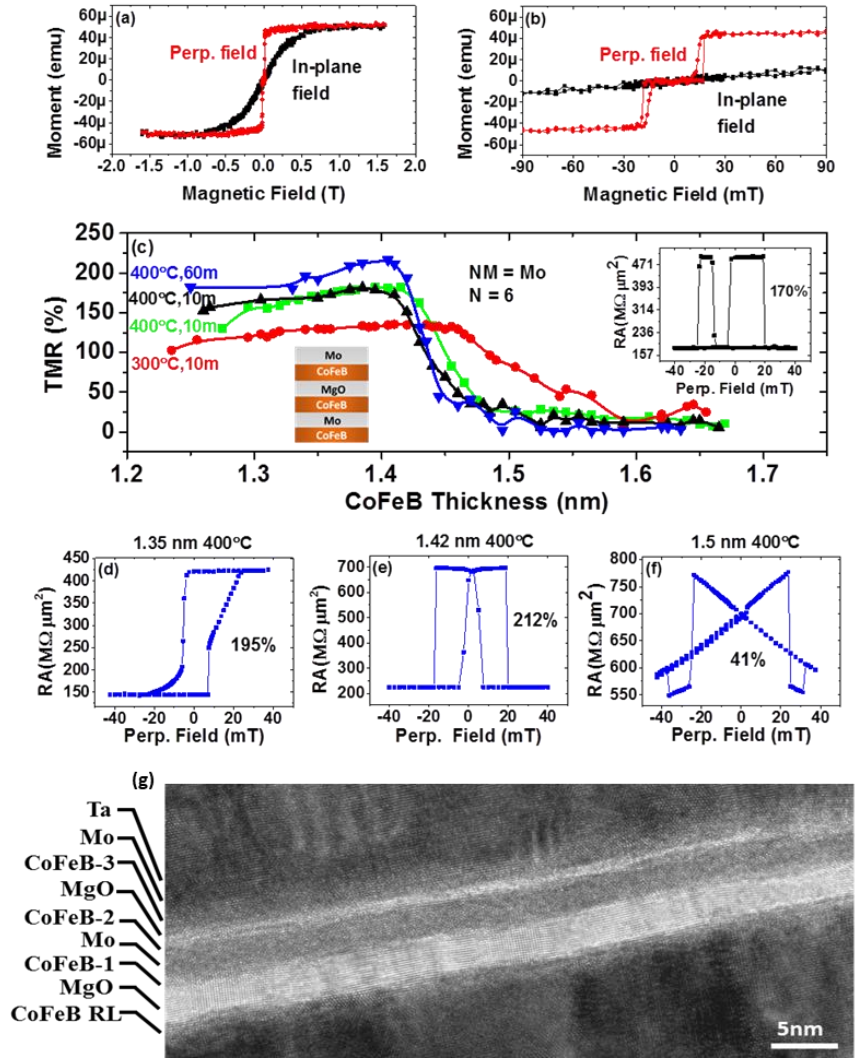

Figure 4. (a) Hysteresis loops of MgO/ CoFeB/Mo(0.9nm)/CoFeB/MgO/CoFeB/Mo measured on different magnetic fields. (b) The same curves at low field-region. (c) CoFeB thickness dependence of the TMR in pMTJs with Mo-MIFLs (Inset: schematic of Mo-MIFL). The samples with Mo-0.9nm coupling layer were successive annealed at 300°C for 10min (red dots), 400°C for 10min (black up-pointing triangles), then 400°C for another 50min (blue down-pointing triangles, where the total annealing time at 400°C is 60min). The green square data points are the TMR values of the samples with the Mo-1nm coupling layer, annealed at 400°C for 10min. The lines are for guiding eyes only. Inset shows the TMR curve of a pMTJ with the Mo-1nm coupling layer. (d-f) Representative TMR curves of three pMTJ after the 400°C annealing for 60min. (g) HRTEM image of the pMTJ with Mo-MIFL.

The microstructure of pMTJ with Mo(0.6nm) -MIFL was investigated by TEM and is presented in Figure 4g. The MgO tunnel barrier exhibits good (001) crystalline structure throughout the specimen. The CoFeB reference layer beneath the MgO tunnel barrier and the first CoFeB in the MIFL show predominantly (001) texture, indicating successful solid-state-epitaxy from the MgO barrier outward during the 400˚C anneal. The successful recrystallization of the CoFeB/MgO/CoFeB complex is critical for the high TMR ratios observed within these pMTJs. The second and the third CoFeB layers in the MIFL, however, are only partially crystallized. The MgO layer in the MIFL is mostly continuous and exhibits partial crystallization from the thermal processing. Note in MIFLs the amorphous second (and third) CoFeB and some pinholes in the MgO could potentially be advantageous, as they may help to reduce the Gilbert damping and the resistance-area product of the devices, respectively. Note for Ta-MIFL with N=6, the maximum TMR is only 180%, again demonstrating the advantage of Mo as a better HM layer.

These results suggest the first CoFeB thickness must be precisely controlled in order to achieve the largest TMR in pMTJs with MIFL. Though the drop of TMR after a certain threshold thickness of the first CoFeB is a common feature observed in all three types of MIFLs in this study, the decay in pMTJs with Mo-MIFL is most pronounced. Obviously, this phenomenon is related to the reduced PMA of the first CoFeB layer when its thickness is getting larger. However the fast drop of TMR with Mo-MIFL cannot solely be explained by PMA, since PMA is usually stronger in pMTJs with Mo compared to those with Ta.[12] The presence of interlayer coupling makes the situation more complicated, which can certainly have a large influence on the switching behavior of the CoFeB layers in the MIFL. In particular, the coupling strength may vary with the thickness of the FM layer, which in the Bruno model is due to the Fabry-Perot-type interferences of the electron wave functions through multiple reflections in FM layers.[46] Meanwhile, the magnetic properties of the CoFeB itself at a given thickness is under constant change (such as crystallization and redistribution of O at the MgO/CoFeB interface) when annealed at different conditions, which will in turn impact the interlayer magnetic coupling. A more detailed study is needed to understand the evolution of interlayer coupling in these pMTJs. The difference of the two sets of Mo-MIFL samples (0.9 nm vs 1 nm) also indicates the range of CoFeB thickness that gives rise to high TMR may be expanded if the AF coupling is enhanced.

To conclude, we have developed pMTJs with MIFL that can potentially simultaneously afford high TMR, strong retention and efficient switching for MRAM cells that are scaled down to small dimensions. Different nonmagnetic materials have been explored as the coupling layer in the MIFL. The TMR of the pMTJ exhibits a strong dependence on the thickness of the first CoFeB layer. Large TMR above 200% has been achieved after 400˚C annealing in pMTJs where three CoFeB layers are incorporated in the MIFL.

**Acknowledgements**

This work was supported in part by Semiconductor Research Corporation through the Logic and Memory Devices program (Global Research Collaboration), by DARPA through the ERI program (FRANC), and by NSF through DMR-1905783. M.A. and A.L. were supported by the REU supplement of NSF ECCS-1554011.

**Data availability**

The data that support the findings of this study are available from the corresponding author upon reasonable request.